\documentclass[]{elsart}
\usepackage{graphicx,natbib,amssymb,psfig}
\journal{New Astronomy}
\begin{document}
\begin{frontmatter}
\title{Is the $\beta$ Pictoris member GJ\,3039AB a physical binary? What the rotation periods tell us. }
\author[INAF]{Sergio Messina\corauthref{cor}},
\corauth[cor]{Corresponding author.}
\ead{sergio.messina@oact.inaf.it}
\author[MO]{Ramon Naves},
\ead{ramonnavesnogues@gmail.com}
\author[ARIES]{Biman J. Medhi}
\ead{biman@aries.res.in}
\address[INAF]{INAF- Catania Astrophysical Observatory, via S.Sofia, 78 I-95123 Catania, Italy} 
\address[MO]{Montcabrer Observatory, C/Jaume Balmes, 24, Cabrils, Spain}
\address[ARIES]{Aryabhatta Research Institute of Observational Sciences, Manora Peak, Nainital 263129, India}
 
\begin{abstract} 
We have carried out a multi-band photometric monitoring of the close visual binary GJ\,3039, 
consisting of a M4 primary and a fainter secondary component, and likely member of the young stellar association $\beta$ Pictoris (24-Myr old). From our analysis we found that both components are photometric  variables \rm and, for the first time, we detected two micro-flare events.
We measured from periodogram analysis of the photometric time series two rotation periods P = 3.355\,d
and P = 0.925\,d, that we could attribute to the brighter GJ\,3039A and the fainter GJ\,3039B components, respectively.
A comparison of these rotation periods with the period distribution of other $\beta$ Pictoris members further supports that
GJ\,3039A is a member of this association.  We find that also GJ\,3039B could be a member, \rm but the infrared magnitude differences between the two components  taken from the literature  and the photometric variability, which is found to be comparable in both stars, suggest that GJ\,3039B could be a foreground star
physically unbound to the primary A component. \rm 
\end{abstract}
\begin{keyword}
Stars: activity - Stars: low-mass  - Stars: rotation - 
Stars: pre main sequence: individual:   GJ3039
\end{keyword}
\end{frontmatter}

\section{Introduction}
Information on the stellar rotation period, especially when combined with  information on 
stellar radius and projected rotational velocity ($v\sin{i}$), provides a powerful tool to better
understand the nature and evolution stage of low-mass stars.
Once the rotation period of a low-mass field star is known, we can estimate its age through the gyro-chronology
technique (see, e.g., Barnes 2007), or,  more generally, we can put some constraint on its age.
The rotation period contributes to assess the membership of a star to a given stellar cluster, association, 
or Moving Group (see, e.g., Messina et al. 2015a). Moreover, when 
a star has a companion in a close orbit,  the rotation period allows us to shed light on the early stage of 
its stellar evolution, specifically on the  primordial disc lifetime.
In fact, the faster rotation that we observe in a close binary with respect to an equal-mass single star
reveals that gravitational effects have enhanced  the disc dispersal, and shortened the disc-locking 
duration (see, e.g., Messina et al. 2014, 2015a, 2016a).
The rotation period combined with the  $v\sin{i}$ can put constraints on the value of the stellar radius
since the inclination of the stellar rotation axis must be $i < 90^{\circ}$. This constraint allows to investigate, for
example, the issue of radius inflation on active stars (see, e.g., Messina et al. 2015c). As well,
the knowledge of the rotation period allows to  interpret more accurately the observed dispersion in the Li distribution among members of  clusters/associations. In fact, rotation has effects on the Li content making fast rotators Li richer with respect to equal-mass slow rotators (see, e.g., Soderblom 1993; Messina et al. 2016b). Finally, the knowledge of the rotation period of low-mass stars allows to model the apparent radial velocity (RV) variations produced by magnetic activity and to remove them in the RV search for planetary companions (Lanza et al. 2011).\\
In the present paper, we present the results of an ongoing study aimed at measuring the rotation periods of 
low-mass stars that are members or candidate members of stellar associations (Messina et al. 2010, 2011). In this paper we present the results of a photometric monitoring of GJ\,3039, a visual binary candidate member of the 24-Myr old $\beta$ Pictoris association. We will use the rotation periods, first measured by us, to asses the membership of the GJ3039 system to this  association and to investigate if the A and B components of this visual binary are effectively bound, forming a physical binary, or if they are distant one from the other and physically unrelated.

\section{Literature information}

GJ\,3039 (NLTT 1741; LP 525-39) was first classified by Stephenson (1986) as a M3 
dwarf from photographic plates.
Weis (1986) reported the following magnitude  V = 12.70\,mag, 
color V$-$I = 2.66\,mag, and a photometric distance d = 13\,pc. It was included in the catalogue of nearby stars with the name GJ\,3039 (Gliese \& Jahreiss 1991).
The first spectroscopic analysis was made by Reid et al. (1995) who classified GJ\,3039 as a M4 star with
 radial velocity RV = 0.9\,km s$^{-1}$, proper motion $\mu_\alpha = 174$\,mas\,yr$^{-1}$,
 $\mu_\delta = -67$\,mas\,yr$^{-1}$, magnitude V = 12.34\,mag, spectroscopic distance d = 11.8\,pc, and 
 space velocity components: U = $-6$\,km\,s$^{-1}$, V = $-$7\,km\,s$^{-1}$, W = $-$3\,km\,s$^{-1}$. 
 GJ\,3039 was associated to the X-ray source RX\,J003234.5+072929 detected by the ROSAT All Sky Survey (Huensch et al. 1999) and included 
 in the catalogue of UV Cet-type flare stars by Gershberg et al. (1999), although,
 for instance, no flares events were ever detected on this star.\\
McCarthy et al. (2001), in their near-infrared high-spatial resolution  
images  (plate scale of 0.15$^{\prime\prime}$/pixel) of GJ\,3039 taken at Keck I, 
discovered the presence of a visual companion at an angular distance $\rho$ = 0.73$^{\prime\prime}$ ($\sim$30\,AU)
and position angle PA = 333$^{\circ}$.92 at the epoch 1997.9. The two components were resolved and the
near-infrared magnitudes of the two components A and B were measured: J$_{\rm A}$ = 9.13\,mag, J$_{\rm B}$ = 9.61\,mag, and K$_{\rm A}$ = 8.31, K$_{\rm B}$ = 8.82 mag. 
Gizis et al. (2002, and references therein) measured an H$\alpha$ EW = 5.59\,\AA,\,\,corresponding to a luminosity ratio L$_\alpha$/L$_{bol}$ = $-$3.63. They also reported  RV = $-$4.6\,km\,s$^{-1}$, U = $-6.9$\,km\,s$^{-1}$, V = $-$11.5\,km\,s$^{-1}$, W = 0.4\,km\,s$^{-1}$.\\
The projected rotational velocity $v\sin{i}$ = 14.7\,km\,s$^{-1}$ was measured by Reiners et al. (2012).
The most recent spectroscopic investigation was carried out by Schlieder et al. (2012a,b) 
to probe its possible membership to the $\beta$ Pictoris association. They measured radial  velocity RV = $-2.9$$\pm$1.1\,km\,s$^{-1}$, projected rotational velocity $v\sin{i}$ = 20$\pm$2\,km\,s$^{-1}$, and a kinematic distance  d = 41.1$\pm$4.4\,pc, a model-derived distance d = 34.1$\pm$17\,pc,\footnote{The earlier estimates of distance, which are all smaller, were derived when GJ3039 was not known to be a visual binary, and therefore, based on the integrated magnitude.} and space velocity components U = $-9.8$\,km\,s$^{-1}$, V = $-$18\,km\,s$^{-1}$, W = $-$5\,km\,s$^{-1}$, proper motions $\mu_\alpha$ = 92\,mas\,yr$^{-1}$,  $\mu_\delta = -55$\,mas\,yr$^{-1}$, Galactic distances  X = $-$9.9$\pm$3\,pc, Y = 21.4$\pm$2.1\,pc, Z = $-$33.7$\pm$3.7\,pc. They also  reported the detection of strong Near-UV and Far-UV emission. From their analysis, they concluded that GJ\,3039 is likely a member of the $\beta$ Pictoris association.

\section{Photometric observations}
To measure the photometric rotation periods of the two components of GJ\,3039 we have explored the public photometric archives, as well as we have carried out our own photometric monitoring.
\subsection{Public archives}
Photometric time series of GJ\,3039  are available in the NSVS (Northern Sky Variability Survey; Wo\'zniak et al. 2004) and in the
 ASAS (All Sky Automated Survey; Pojmanski 1997) public archives.
From the NSVS we retrieved 114 measurements  spanning the time interval from  November 1999 to February 2000, whereas
from the ASAS archive we could retrieve 235 V-band photometric measurements collected from 2001 to 2009. 
As discussed in the next Section, the periodogram analysis of NSVS data provided one possible rotation period, but to be confirmed by additional observations. The ASAS data did not provide any significant periodicity. For these reasons,
we decided to plan our own photometric monitoring.

\subsection{Montcabrer observations}
	We observed GJ\,3039 at Montcabrer Observatory in Spain (MO, $41^{\circ}\,31^{\prime}\,11^{\prime\prime}$\,N, 
	$02^{\circ}\,23^{\prime}\,39^{\prime\prime}$E\,, 100m a.s.l.) in 2014, from October 17 to November 
	17 for a total of 16 nights. Observations were performed with a  30-cm f/6 Meade LX-200 telescope equipped with a SBIG ST-8 CCD camera,
	 AO-8T active optics, and Bessel BVRI filters. The telescope has a corrected  26.3$^{\prime}\times17.5^{\prime}$ field of view (FoV) and a 
	plate scale of 1.03$^{\prime\prime}$/pixel. On each night, the target was observed up to six consecutive hours, using 120-s integration,  
	allowing us to collect a total of 1750 frames in the R, 60 in the V, and 60 in the I filters. In all frames, the A and B component of GJ\,3039 were spatially unresolved.
	The data reduction was performed using the IRAF\footnote{IRAF is distributed by the National Optical Astronomy Observatory, which 
	is operated by the association of the Universities for Research in Astronomy, inc. (AURA) under cooperative agreement with the National 
	Science Foundation.} tasks within DAOPHOT. After dark and bias subtraction, and flat field correction of the science frames,  we extracted 
	the V, R, and I magnitudes time series of GJ\,3039 and other stars in the vicinity to search for suitable comparison stars. 
	These magnitude time series were obtained using the aperture photometry, and were subsequently cleaned by applying a 3$\sigma$ threshold
	 to remove outliers. We could identify one bright (HD\,2934) and four fainter stars whose light curves were found to be very stable ($\sigma_R < $ 0.008 mag)
	 during the whole observation run and then well suited to be used as comparison and check stars.
	Differential magnitudes of GJ\,3039 were obtained with respect to the star HD\,2934 (RA = 00:32:43.49, DEC = +07:35:24.98, J2000.0, V = 9.26\,mag, R = 8.8\,mag, I = 8.4\,mag).
	For the subsequent period search, consecutive magnitudes collected within a time interval of 30 min (about 6--7 consecutive measurements) were averaged. That allowed us 
	to have a better estimate of the achieved photometric precisions, which turned to be: $\sigma_{\rm V}$ = 0.023\,mag, $\sigma_{\rm R}$ = 0.006\,mag, 
	$\sigma_{\rm I}$ = 0.003\,mag.
	
\subsection{ARIES observations}
 We could observe GJ\,3039 also at the Aryabhatta Research Institute of Observational Sciences (ARIES,  
 $29^{\circ}\,22^{\prime}\,49^{\prime\prime}$\,N, $79^{\circ}\,27^{\prime}\,47^{\prime\prime}$\,E,  Manora Peak,  Nainital, India) 
 in three nights, November 27, 28 and December 26, 2014. More observation nights were
 prevented by bad weather conditions. Photometric observation were collected with a  104-cm f/13 telescope equipped with a 2K$\times$2K CCD camera providing a corrected 
 13$^{\prime}\times13^{\prime}$ FoV, and a 
	plate scale of 0.366$^{\prime\prime}$/pixel. We could collect a total of 81 frames in the V filter using 60-sec (90-sec in poor seeing conditions) integration time. Data reduction was 
	performed as for the Montcabrer Observatory data. The average photometric accuracy that we could achieve was $\sigma_V$ = 0.003 mag. Owing to the smaller FoV, we could use the four fainter check
	stars to build an ensemble comparison star and to get the V-band  differential light curve of GJ\,3039. In the subsequent period search analysis, we added a magnitude off-set to make V magnitude time series comparable to
	the R magnitude time series collected at Montcabrer Observatory.

In both observation runs, at MO and at ARIES, the small 0.$^{\prime\prime}$79 angular separation between the components of GJ\,3039 did not allow us to spatially resolve them.
Therefore, the observed magnitudes refer to the unresolved system.

\section{Flare activity}
GJ\,3039 was included  in Gershberg et al. (1999) catalogue that lists  UV Cet stars and related objects. Actually, detection of flares in this system was never reported in the literature. 
A very interesting result of our photometric monitoring was the discovery that GJ\,3039 during two nights exhibited  such sudden brightenings that can be interpreted as flare events.
These events, detected at MO on October 17 (HJD = 2456951.50 ) and November 16 2014 (HJD = 2456978.47) (Fig.\,1), exhibit the typical shape of a micro-flare with a quick magnitude brightening (larger than $\Delta$R = 0.05\,mag the first and $\Delta$R = 0.04\,mag the last) followed by a less rapid exponential-type decrease to the quiet state, and lasting not more than 30--45 min.
The brightest magnitude is significantly different, at 4.5$\sigma$ level, from the quiet state. However, we have observations in only one band and other events were not
detected. Therefore, we consider those as 'possible' micro-flares.
We note that both flares were detected in correspondence of the phases of light curve minimum, i.e. when spots are best visible, in both components: $\phi$ = 0.60 and 0.78 for component A, and $\phi$ = 0.38 and 0.39 for component B (see next Section).

\section{Rotation period search}
We used the Lomb-Scargle (LS; Scargle 1982) and CLEAN (Roberts et al. 1987) periodogram
analyses to search for the rotation periods of the components of GJ\,3039.\\
The photometric precision of ASAS archive data  of GJ\,3039 is quite low ($\sigma_{\rm V}$ = 0.030 mag). From the periodogram analysis of ASAS 
data we did not infer any rotation period with False Alarm Probanbility FAP $<$ 1\%. The False Alarm Probability, that is the probability that a power peak of that height simply
arises from Gaussian noise in the data,  was estimated
using a Monte-Carlo method, i.e., by generating 1000 artificial light
curves obtained from the real one, keeping the date but scrambling
the magnitude values (see, e.g., Herbst et al. 2000).\\
The photometric precision of NSVS archive data of GJ\,3039 is   better ($\sigma_{\rm V}$ = 0.014 mag). From the periodogram analysis of NSVS 
data, we inferred one power peak at P = 0.909$\pm$0.020\,d (see Fig.\,2) with a confidence level of  95\%.
However,  this level is smaller than the typical 99\% adopted 
in our previous similar studies (see, e.g., Messina et al. 2011). Therefore, we planned our own observations to confirm it and to search 
for the rotation period of the other component.\\
The LS analysis of our new photometry revealed a number of highly significant (FAP $<$1\%)
power peaks in the periodogram of the R-magnitude time series.  The two major power peaks correspond to the rotation periods  P = 3.355$\pm$0.008d and
P = 0.925$\pm$0.005d.  Most secondary power peaks in the LS periodogram (top-middle panel of Fig.\,3) 
are related to the beating between the primary periods and the 1-d
sampling imposed by the Earth rotation and the fixed longitude of the observation site. The CLEAN algorithm effectively removed such beat periods leaving 
only two significant  peaks (top-right panel of Fig.\,3) at the same rotation periods detected by LS.
The normalized power corresponding to a FAP = 1\%
was P$_N$ = 11 (dashed horizontal line).
The uncertainties of the rotation periods are computed following the prescription of Lamm et al. (2004).

As mentioned, the small 0.$^{\prime\prime}$79 angular separation between the components of GJ\,3039 did not allow us to spatially resolve them,
therefore, we obtained integrated magnitudes. However, since the components have comparable brightness (see Sect.\,2) we expect that both 
components can significantly contribute
to the observed variability, leaving signs of their rotation periods in the time series light modulation.
Therefore, once we identified the primary period P = 3.355\,d, we filtered it out from the magnitude time series and recomputed
the periodograms. As shown in Fig.\,4, we recovered the other highly significant period P = 0.925$\pm$0.005d, which is in agreement, within the uncertainties, with the rotation period inferred from the NSVS data. 

Therefore, we conclude that both components of GJ\,3039 are variable and their rotation periods are P = 3.355$\pm$0.008d and P = 0.925$\pm$0.005d. We note that in both components the differential color variation $\Delta$(V$-$I) is positively correlated to the differential $\Delta$R magnitude variation: the star is redder when it is fainter. This is the kind of correlation observed in spotted star, when the photometric variability arises from the presence of spotted regions, either cooler or warmer than the unperturbed photosphere, unevenly distributed across the stellar longitudes.

\section{Discussion}

The brightest observed magnitude for the unresolved system is V = 12.68\,mag, as inferred from the long ASAS time series. In the case that both components have same
brightness, the correction for duplicity is $\Delta$V = 0.72\,mag. However, from the near-infrared magnitude measurements by 
McCarthy et al. (2001), we infer that one component should be fainter than the other
 and the correction to apply should be $\Delta$V $<$ 0.72\,mag.
In the following analysis we use a weighted average for the projected rotational velocity $<v \sin{i}>$ = 17.0\,km\,s$^{-1}$.
We use
the bolometric corrections BC$_{\rm V}$ = $-$2.43\,mag and effective temperatures T = 3160\,K 
corresponding to a M4 spectral type, according to Pecaut \& Mamajek (2013). Combining projected rotational velocity, 
rotation period, and stellar radius, we can constraint the value for the inclination $i$ of the rotation axis.\\
The brighter M4 component of the system may have either P = 3.355\,d or P = 0.925\,d as rotation period.\\
Let us explore the first possibility,  that is P = 3.355\,d is the rotation period of the brighter A component. 
In order to have $\sin{i} \le 1$ we need that R$_{\rm A} \ge 1.1$\,R$_\odot$. Such values of stellar radii can be obtained if the bolometric magnitude is M$_{\rm bol_A} \le 7.1$\,mag. Again, such range of values can be obtained imposing that V$_{\rm A}$$-$5$\log$(d) $\le$ 4.53. Since the observed magnitude must be 12.68 $\le$ V $\le$ 13.4\,mag, consequently we get  43 $\le$ d $\le$ 59\,pc. This range of distance partly overlaps, within the uncertainties, with the kinematic distance d = 41.1$\pm$4.4\,pc inferred by Schlieder et al. (2012). Therefore, it is reasonable to attribute the rotation period   P = 3.355\,d to the brighter A component of GJ\,3039.
For instance, either for fainter magnitudes or larger distances (within the mentioned ranges) we will get lower values of the inclination of the  rotational axis.\\
The second possibility is that  P = 0.925\,d is the rotation period of the brighter A component. Following the same reasoning,
in order to be  0.4 $ \le \sin{i} \le 1$ we need   0.3 $\le$ R$_{\rm A} \le 0.7$\,R$_\odot$. We note that very low values of the inclination ($\sin{i} \le 0.4$) in combination with the light dilution effect owing to the unresolved photometry,  would prevent any light rotational modulation, which is not our case. The stellar radius is within such range of values  if the bolometric magnitude is 8.4 $\le$ M$_{\rm bol_A} \le 9.9$\,mag. Again, such range of values can be obtained imposing that 5.8 $\le$ V$_{\rm A}$$-$5$\log$(d) $\le$ 7.3. This constraint can be respected if we use either V $\ge$ 13.8\,mag or 12 $\le$ d $\le$ 33\,pc. The first is a magnitude too faint even if the correction for duplicity is applied, and the second is outside the  kinematic distance d = 41.1$\pm$4.4\,pc inferred by Schlieder et al. (2012).\\
On the basis on this considerations, we can state that GJ\,3039A has a rotation period P$_{\rm A}$ = 3.355\,d, and an inclination of its rotation axis $i$$\sim$90$^{\circ}$, and GJ\,3039B has a rotation period P$_{\rm B}$ = 0.925d.\\

Once we have attributed the rotation periods to the respective components, we can try to investigate if the two components are or are not physically bound. In fact, for the secondary component of this system, as far as we have found in the literature, we have no measurements of distance, space velocity components, RV, proper motion. Therefore, it may be a physical companion, but also either a foreground or a background star. To investigate this issue we can use the available near-infrared photometry.\\
 
For this visual binary we have accurate integrated J and H magnitudes from 2MASS (Cutri et al. 2003), J$_{\rm AB}$ = 8.839$\pm$0.024\,mag 
and K$_{\rm AB}$ = 7.508$\pm$0.018\,mag collected around 2000.5, and less accurate values for the resolved components by McCarthy et al. (2001), J$_{\rm A}$ = 9.13$\pm$0.15\,mag, J$_{\rm B}$ = 9.61$\pm$0.15\,mag, and K$_{\rm A}$ = 8.31$\pm$0.15\,mag, K$_{\rm B}$ = 8.82$\pm$0.15\,mag collected around 1997.9.
The circumstance that the photometry is collected at two epochs, about 2.6 yr distant from each other, may imply some effects of the long-term variability on the measured magnitudes. This is, however, of the order of a few hundredths of magnitude,
 as inferred from the 9-yr long ASAS time series, and significantly smaller than the quoted photometric uncertainties.
\\

\rm

Assuming that A and B form a physical binary and, therefore, have the same distance from the Sun, the flux ratios between the two components can be derived in three different ways, using the un-resolved 2MASS photometry together with the resolved photometry from McCarthy et al. (2001) or from the latter alone, which however has larger uncertainties.\\
\it Case a: \rm  from the apparent magnitude difference J$_{\rm AB}$ $-$ J$_A$ = $-$0.73$\pm$0.15\,mag we derive (F$_A$/F$_B$)$_J$ = 1.04$\pm$0.25
and from  K$_{\rm AB}$ $-$ K$_A$ = $-$0.8$\pm$0.15\,mag we derive (F$_A$/F$_B$)$_K$ = 0.91$\pm$0.25.
These flux ratios lead to the magnitude differences $\Delta$J = $-$0.04$\pm$0.20\,mag and  $\Delta$K = $-$0.10$\pm$0.20\,mag.
Considering the small magnitude difference,  we infer that both components have same M4 spectral type.\\
\it Case b: \rm from the apparent magnitude difference J$_{\rm AB}$ $-$ J$_B$ = $-$1.21$\pm$0.15 mag we derive (F$_A$/F$_B$)$_J$ = 2.05$\pm$0.25
and from  K$_{\rm AB}$ $-$ K$_B$ = $-$1.31$\pm$0.15 mag we derive (F$_A$/F$_B$)$_K$ = 2.35$\pm$0.25.
These flux ratios lead to the magnitude differences $\Delta$J = $-$0.78$\pm$0.20\,mag and  $\Delta$K = $-$0.93$\pm$0.20\,mag.
A comparison with model flux ratios for a 24\,Myr isochrone by Siess et al. (2000) indicates that
 the secondary must have a M5.5 spectral type, to which corresponds a model V magnitude difference $\Delta$V = 2.0$\pm$0.5\,mag.\\
\it Case c:  \rm finally, if we consider only  the resolved magnitudes from McCarthy et al. (2001), we have $\Delta$J = $-$0.45$\pm$0.20\,mag and $\Delta$K = $-$0.51$\pm$0.20\,mag, from which we derive (F$_A$/F$_B$)$_J$ = 1.5$\pm$0.30 and (F$_A$/F$_B$)$_K$ = 1.5$\pm$0.30. A comparison with the Siess et al. models indicates that, in this case,  the secondary component must have a spectral type around M5.3, to which corresponds a model V magnitude difference $\Delta$V = 1.5$\pm$0.5\,mag.\\

In GJ\,3039 both components have comparable level of photometric variability that allowed us to measure the rotation periods of both components. Considering the light dilution effect that, in the case of unresolved systems, reduces the amplitude of the photometric variability, we infer that the A and B components of GJ\,3039 must have similar magnitudes.

Therefore, if we assume that both components have same distance and form a bound system,
then only case a) is reasonable. In this hypothesis, the disagreement with the other two cases likely arises from an incorrect measurement of the unresolved magnitude for the B component.

On the contrary, if the secondary component has a later spectral type (M5.5 as in case b or M5.3 as in case c), to compensate the
magnitude difference and have the B component of similar brightness and similarly detectable photometric variability, 
its distance to the Sun must be closer than the A component, i.e., it must be a foreground star.
In this hypothesis, the disagreement with case a) arises from the wrong assumption of the same distance for both components, and no erroneous magnitude measurement must be invoked.  Of course, considering the proximity of this system to the Sun and the very small angular separation, the probability that both components are significantly distant from each other, and unbound is very low.

\rm

As shown in Fig.\,5, a comparison with the period distribution of the $\beta$ Pictoris members (Messina et al. 2016a) shows that GJ\,3039A has a rotation period that very well fits into the period distribution of single members and wide components of binary systems (separation $>$ 60\,AU)\footnote{Since GJ\,3039A is seen almost equator on, assuming co-planarity between the orbital and equatorial planes, the projected separation of 30\,AU likely underestimates the effective separation, making GJ\,3039A and B components of a wide binary.}.
Therefore, our rotation period determination further supports its membership to the $\beta$ Pictoris association, as suggested by Schlieder et al. (2012).\\
GJ\,3039B to be member and its rotation period to fit into the same period distribution should have a V$-$K color $\ge$ 5.8\,mag, i.e., a spectral type equal to or later than M5. If this is the case, GJ\,3039B could be a $\beta$ Pictoris member. However, its flux to be comparable to that of the A component and, therefore, its photometric variability to be detectable, GJ\,3039B should be  a foreground star at a distance smaller than GJ\,3039A, and physically unbound.

Finally, in the case the B component is itself an undetected close binary with a much fainter companion, then it may be well
physically bound to the A component and its rotation period would be expected to be significantly shorter than that of the A component, as found in other triple systems (see Messina et al. 2016a).

\rm

\section{Conclusions}
We have carried out a multi-band photometric monitoring of the close visual binary GJ\,3039. Observations were carried out almost contemporary in two different observatories. In both observatories, we could not  resolve spatially the two components. We found that both components are photometric variables and, for the first time, we detected  two micro-flare events in the R filter that confirm the classification of this system in the UV Cet-type class. We measured two rotation periods P = 3.355\,d and P = 0.925\,d, that we could confidently attribute to the GJ\,3039A and the GJ\,3039B components, respectively. The shorter period is also in agreement with the period we could measure from archived NSVS observations. 
The rotation period P = 3.355\,d of GJ\,3039A well fits into the period distribution of other $\beta$ Pictoris members,
which gives additional support to the membership assignment based on other indicators.  The rotation period P = 0.925\,d of GJ\,3039 B could fit into the same distribution if we assign to this component a spectral type equal to or later than M5, or if we assume that GJ\,3039 B is a M4 star, physically bound to the A component, but with an undiscovered very close fainter companion that explains its much faster rotation rate. 
Indications that  GJ\,3039 B may have a M5 or later spectral type also come when we 
 combine the near infrared colors of both components measured by McCarthy et al. (2001) with the 2MASS integrated magnitudes, and compare the measured with the model flux ratios. In this case, however, GJ\,3039B to significantly contribute to the observed variability  must be closer to the Sun than the A component. This circumstance  would  imply that the A and B components are still members, but physically unbound.

\rm 
\section*{Acknowledgments}  
The extensive use of the SIMBAD and ADS databases operated by the CDS centre, Strasbourg, France, is gratefully acknowledged. We thank the anonymous Referee who helped us to improve the quality of the paper.

\clearpage
\noindent
\begin{figure*}
\begin{centering}
\includegraphics[width=100mm,height=130mm,angle=90,trim= 00 0 0 100]{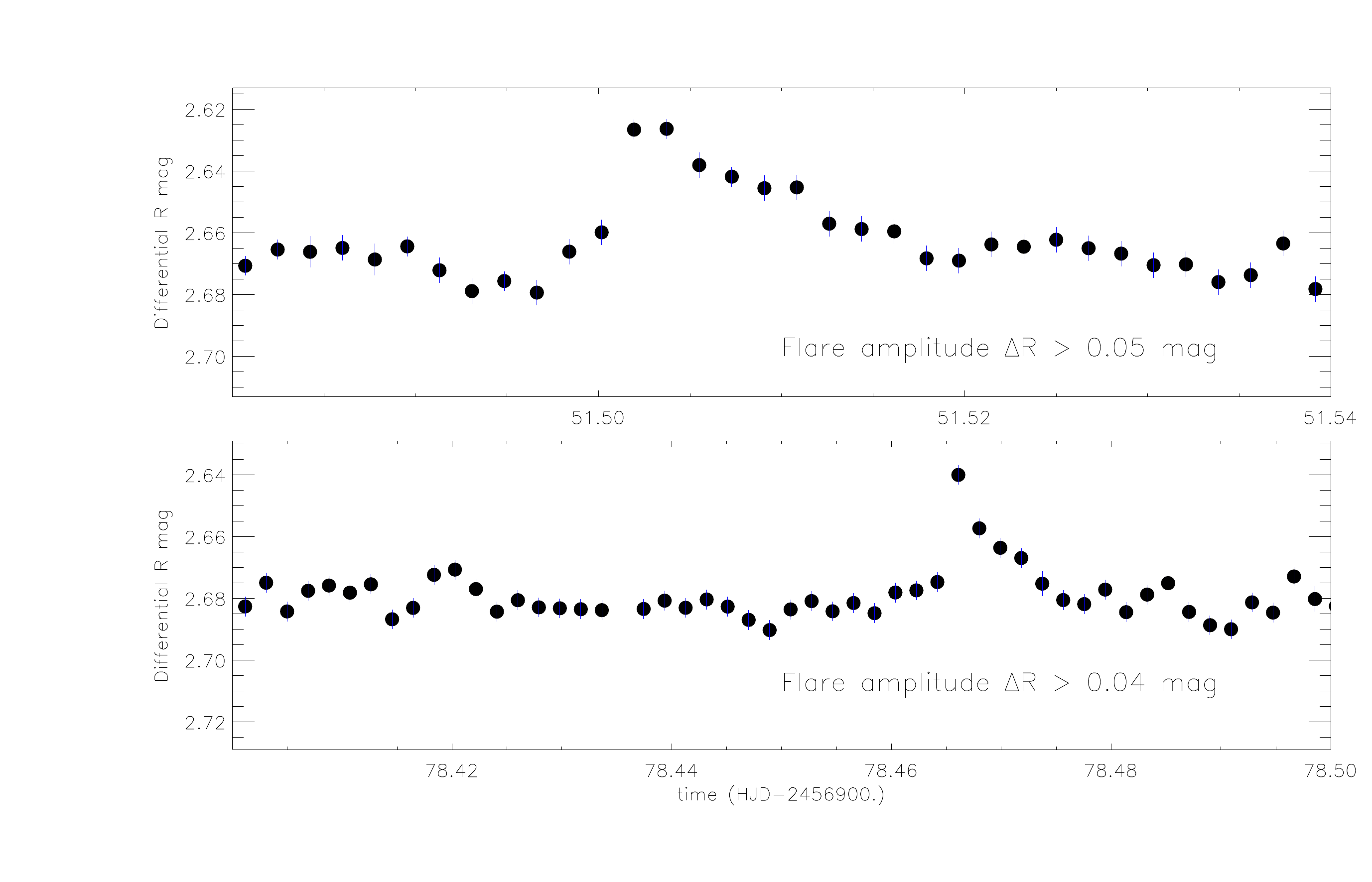}
\vspace{0cm}
\caption{Flare-like events detected in the R filter during two different nights at MO.}
\end{centering}
\label{flare}
\end{figure*}

\begin{figure*}
\begin{centering}
\includegraphics[width=100mm,height=130mm,angle=90,trim= 0 0 0 100]{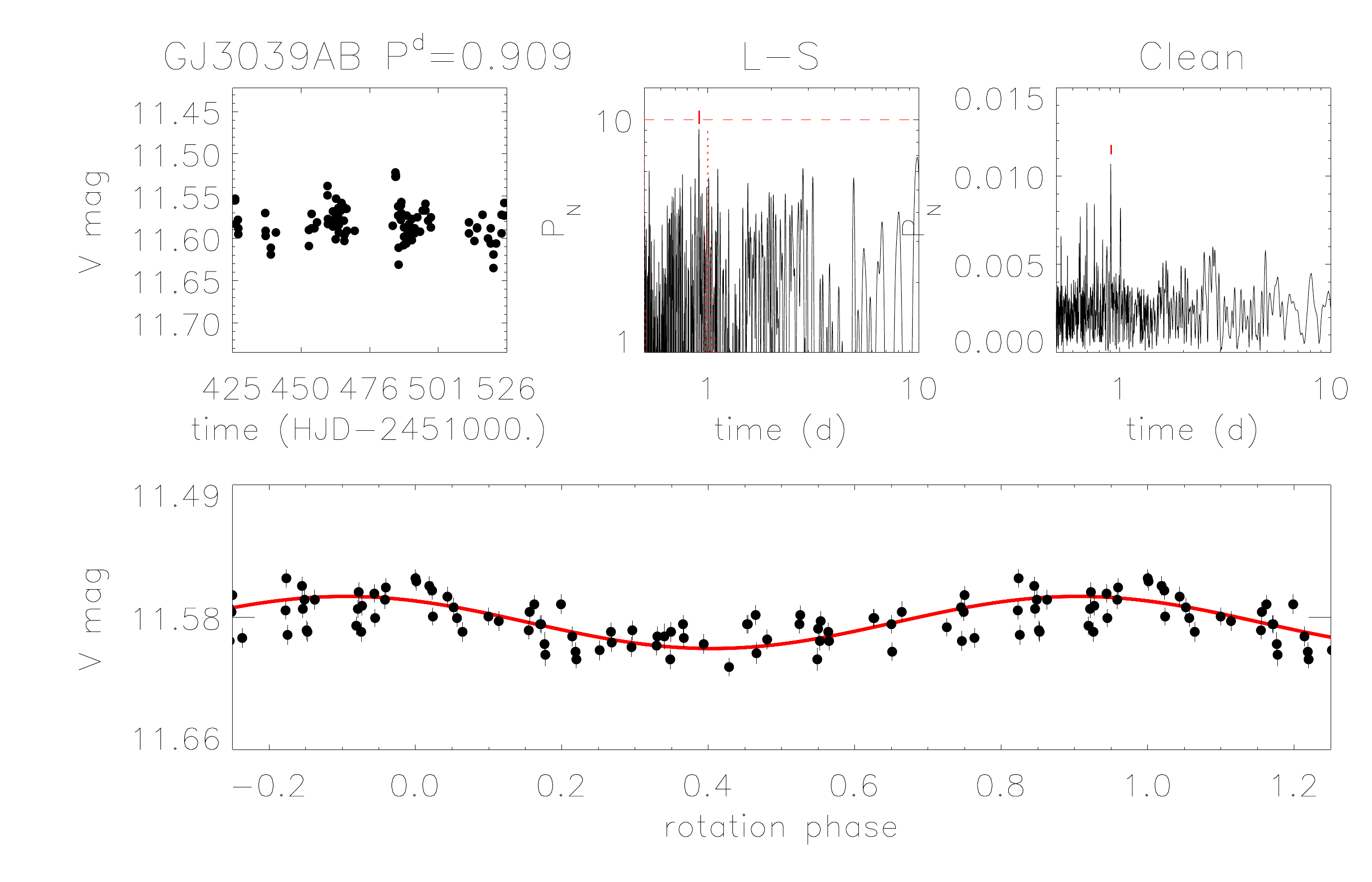}\\
\vspace{1cm}
\caption{\it top panels: \rm NSVS magnitude time series of GJ\,3039; LS periodogram with indication of the 99\% condidence level (horizontal dashed line), and CLEAN periodogram. 
\it bottom panel: \rm light curve phased with the  P = 0.909\,d rotation period and with overplotted (solid line) a sinusoidal fit with an amplitude of $\Delta$V = 0.03\,mag.}
\label{NSVS}
\end{centering}
\end{figure*}

\begin{figure*}
\begin{centering}
\includegraphics[width=100mm,height=130mm,angle=90,trim= 0 0 0 100]{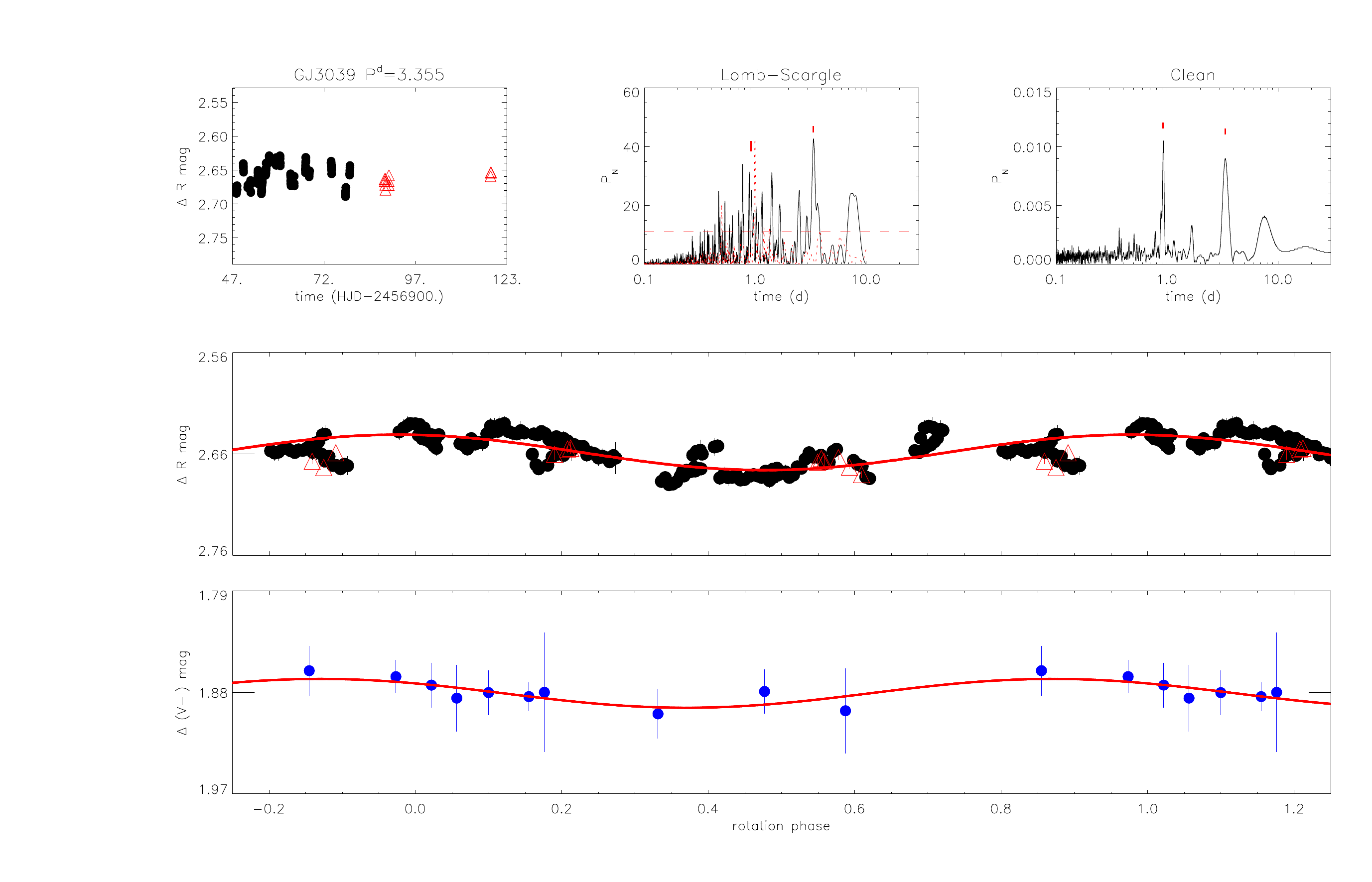}
\vspace{0cm}
\caption{\it top panels: \rm (Left) Differential R-magnitude time series. Black bullets are MO data, red triangles are ARIES data.
(Middle) LS  and (right) CLEAN  periodograms. The dotted line is the spectral window, and the horizontal dashed line indicates the power level corresponding to 99\% confidence level.
\it middle panel: \rm light curve phased with the P = 3.355\,d rotation period, with overplotted a sinusoidal fit with amplitude $\Delta$R =  0.045 mag. \it bottom panel: \rm V$-$I differential color curve phased with the P = 3.355\,d rotation period.}
\end{centering}
\label{GJ3039}
\end{figure*}

\begin{figure*}
\begin{centering}
\includegraphics[width=100mm,height=130mm,angle=90,trim= 00 0 0 100]{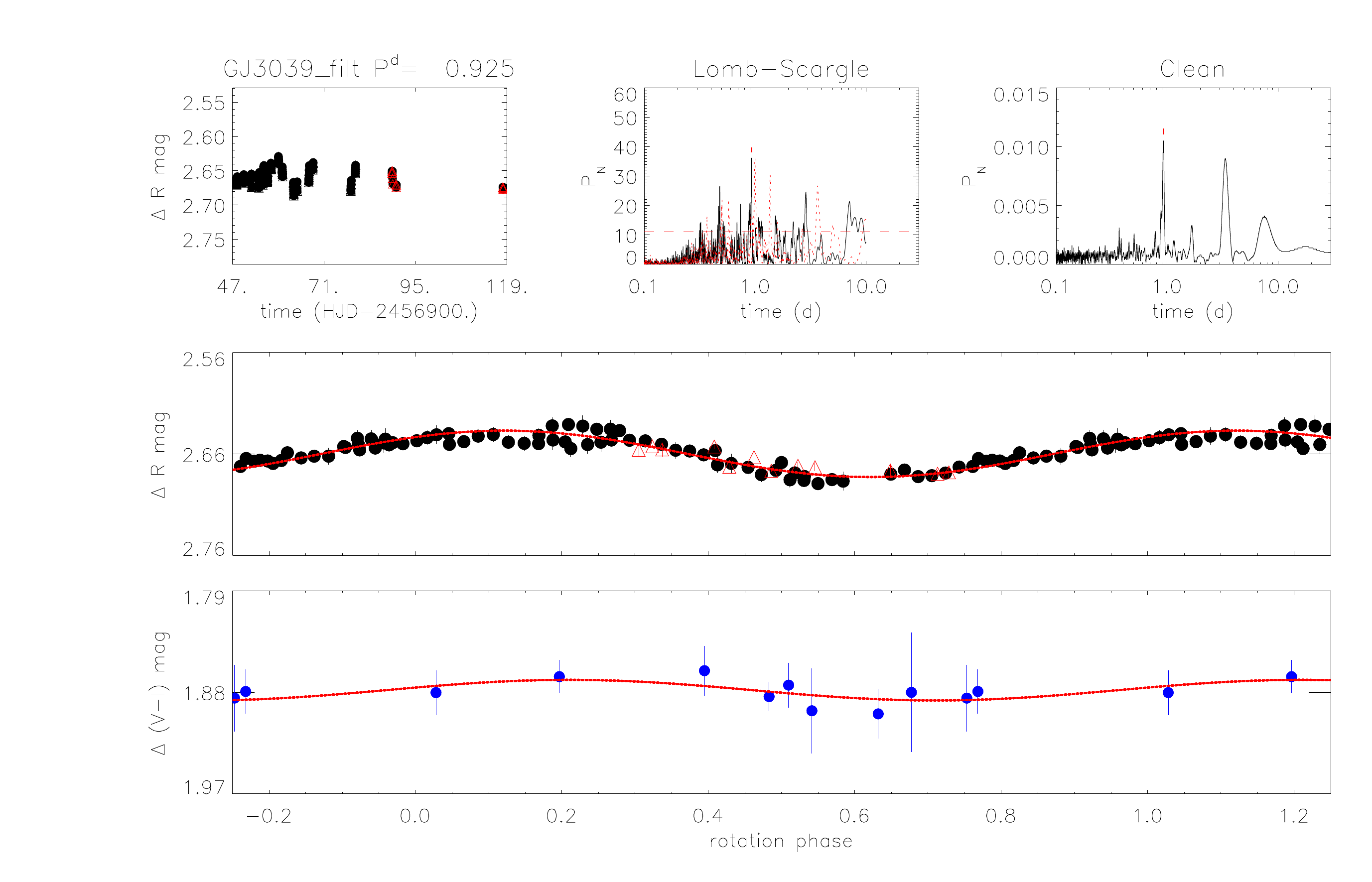}
\vspace{0cm}
	\caption{The same as in Fig\,3, but after filtering the period P = 3.355\,d from the magnitude time series. The CLEAN periodogram is the same as in Fig.\,3. The solid line
in the bottom panel is a sinusoidal fit with period P = 0.925\,d and amplitude $\Delta$R =  0.035 mag.}
\end{centering}
\label{GJ3039_filt}
\end{figure*}

\begin{figure*}
\begin{centering}
\includegraphics[width=100mm,height=130mm,angle=90,trim= 00 0 0 100]{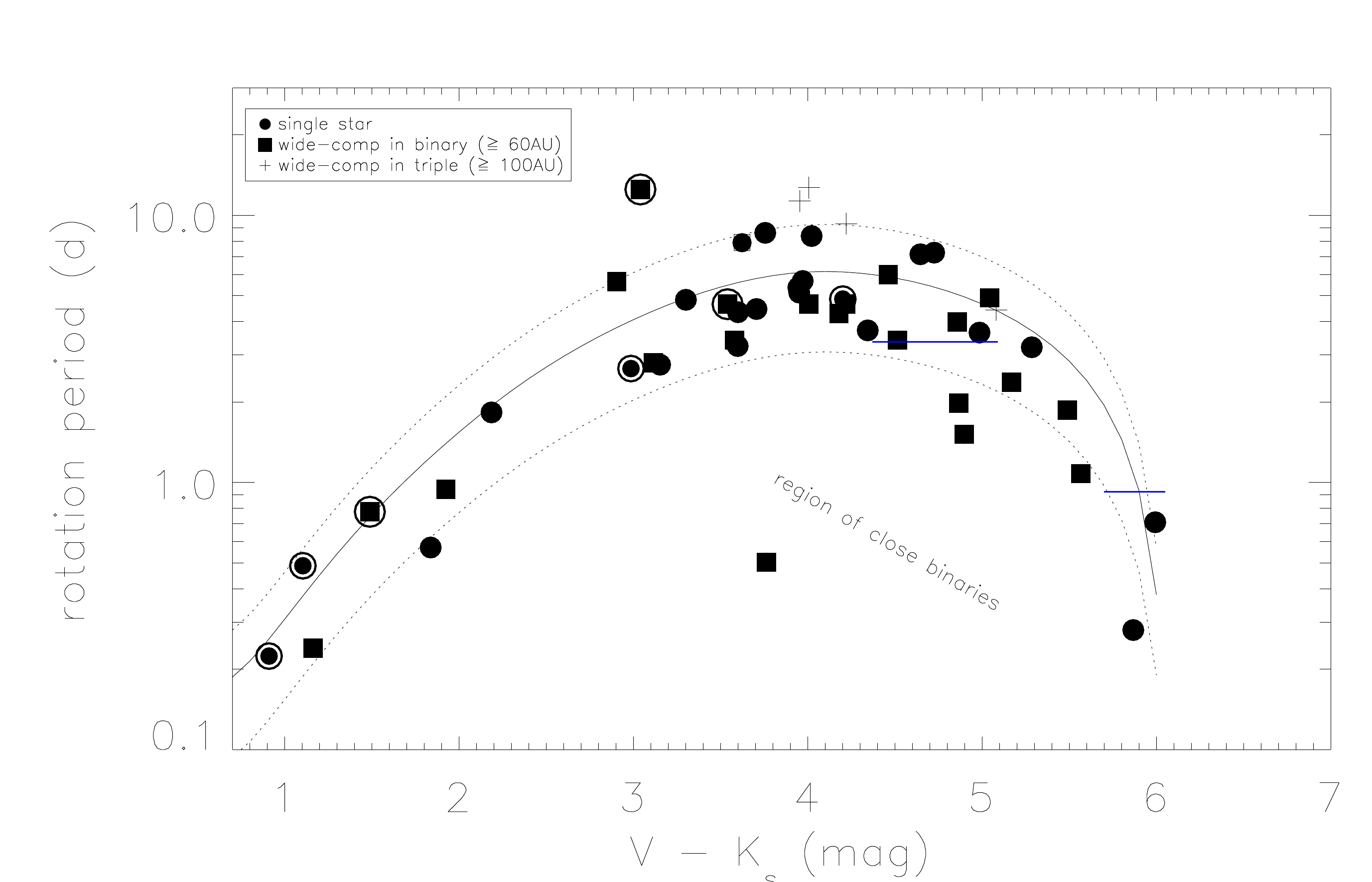}
\vspace{0cm}
	\caption{Distribution of rotation periods of single stars and components of wide binary/triple systems in the $\beta$ Pictoris association (Messina et al. 2016a). Circled symbols are stars hosting a debris disc. The solid black line is a polynonial
	fit to the period distribution (dotted lines denote the 1$\sigma$ dispersion). The solid blue lines indicate the positions of GJ\,3039A and GJ3039B, according to the assigned range of V magnitudes.}
\end{centering}
\label{GJ3039_filt}
\end{figure*}


\begin{thebibliography} {}

\bibitem[Barnes(2007)]{Barnes07}Barnes, S.A. 2007, ApJ, 669, 1167
\bibitem[Cutri et al.(2003)]{Cutri03} Cutri, R. M., Skrutskie, M. F., van Dyk, S. et al. 2003, 2MASS All-Sky Catalog of Point Sources, VizieR On-line Data Catalog: II/246
\bibitem[[Gershberg et al.(1999)]{Gershberg99} Gershberg, R.E., Katsova, M.M., Lovkaya, M.N., Terebizh, A.V. \& Shakhovskaya N.I. 1999, A\&AS, 139, 555 
\bibitem[[Gizis et al.(2002)]{Gizis02}Gizis, J.E., Reid, I.N. \& Hawley, S.L. 2002, AJ, 123, 3356 
\bibitem[Gliese \& Jahreiss]{Gliese91} Gliese, W. \& Jahreiss, H. 1991,  Nearby Stars, Preliminary 3rd Version, 0
\bibitem[Herbst et al.(2002)]{Herbst02} Herbst, W., Bailer-Jones, C. A. L., Mundt, R., Meisenheimer, K., \&  Wackermann, R. 2002, A\&A, 396, 513
\bibitem[Huensch et al.(1999]{Huensch 99}Huensch, M., Schmitt, J.H.M.M., Sterzik, M.F. \& Voges, W. 1999, A\&AS, 135, 319 
\bibitem[Lamm et al.(2004)]{Lamm04} Lamm, M. H., Bailer-Jones, C. A. L., Mundt, R., Herbst, W., \& Scholz, A. 2004, A\&A, 417, 557
\bibitem[Lanza et al.(2011)]{Lanza11}Lanza, A. F., Boisse, I., Bouchy, F., Bonomo, A. S., \& Moutou, C. 2011, A\&A,  533, A44
\bibitem[McCarthy et al.(2001)]{McCarthy01} McCarthy, C., Zuckerman, B. \& Becklin, E.E. 2001, AJ, 121, 3259 
\bibitem[Messina et al.(2010)]{Messina10}Messina, S., Desidera, S., Turatto, M., Lanzafame, A. C., \& Guinan, E. F. 2010, A\&A, 520, A15
\bibitem[Messina et al.(2011)]{Messina11}Messina, S.; Desidera, S.; Lanzafame, A. C.; Turatto, M.; Guinan, E. F. 2011, A\&A, 532, A10
\bibitem[Messina et al.(2014)]{Messina14} Messina, S., Monard, B., Biazzo, K., Melo, C. H. F., \& Frasca, A. 2014, A\&A, 570, A19
\bibitem[Messina et al.(2015a)]{Messina15a} Messina, S., Millward, M., \& Bradstreet, D.H. 2015a, NewAstronomy, 37, 105
\bibitem[Messina et al.(2015b)]{Messina15b} Messina, S., Hentunen, V.-P., \& Zambelli, R. 2015b, IBVS, 6145
\bibitem[Messina et al.(2015c)]{Messina15c}Messina, S., Muro Serrano, M., Artemenko, S., et al. 2015c, Ap\&SS, 360, 51 
\bibitem[Messina et al.(2016a)]{Messina16a}Messina, S.,  Monard, B., Worters, H.L., et al. 2016, NewAstronomy, 42, 29
\bibitem[Messina et al.(2016a)]{Messina16a}Messina, S., Desidera, S., Lanzafame, A.C., et al. 2016a, in preparation
\bibitem[Messina et al.(2016b)]{Messina16b}Messina, S.,  Millward, M., Desidera, S., et al. 2016b, submitted to A\&A
\bibitem[Pecaut \& Mamajek(2013)]{Pecaut13}Pecaut, M. J. \& Mamajek, E. E.  2013, ApJS, 208, 9
\bibitem[Pojmanski(1997)]{Pojmanski97} Pojmanski, G.  1997, AcA, 47, 467
\bibitem[Reid  et al.(1995)]{Reid95} Reid, I.N., Hawley, S.L., \& Gizis, J.E. 1995, AJ, 110, 1838 
\bibitem[Reiners et al.(2012)]{Reiners12} Reiners, A., Joshi, N. \& Goldman, B.  2012, AJ, 143, 80 
\bibitem[Roberts et al.(1987)]{Roberts87}     Roberts, D. H., Lehar, J., \& Dreher, J. W. 1987, AJ, 93, 968
\bibitem[Scargle(1982)]{Scargle82} Scargle, J. D. 1982, ApJ, 263, 835 
\bibitem[Schlieder et al.(2012a)]{Schlieder12a} Schlieder ,J.E., Lepine, S. \& Simon M. 2012, AJ, 143, 80
\bibitem[Schlieder et al.(2012b)]{Schlieder12b}  Schlieder ,J.E., Lepine, S. \& Simon M. 2012, AJ, 144, 109 
\bibitem[Siess et al.(2000)]{Siess00}Siess L., Dufour E., Forestini M. 2000, A\&A, 358, 593
\bibitem[Soderblom et al.(1993)]{Soderblom93}Soderblom, D. R., Jones, B.F.,  Balachandran, S, et al. 1993, AJ, 106, 1059
\bibitem[Stephenson(1986)]{Stephenson86} Stephenson, C.B. 1986, ApJ, 300, 779 
\bibitem[Weis(1986)]{Weis86} Weis, E. W. 1986, AJ, 91, 626
\bibitem[Wo\'zniak et al.(2004)]{Wozniak,04}Wo\'zniak, P. R., Vestrand, W. T., Akerlof, C. W., et al. 2004, AJ, 127, 2436 
\end{thebibliography}
\end{document}